\documentclass[12pt]{article}
\newcommand{\re}[1]{(\ref{#1})}

\newcommand{\g}{\lambda d}

\newcommand{\up}{\uparrow}

\newcommand{\dn}{\downarrow}

\newcommand {\dis}{\displaystyle}

\newcommand{\beg}{\begin{equation}}
\newcommand{\en}{\end{equation}}

\newcommand{\eps}{\epsilon}
\newcommand{\lam}{\lambda}

\usepackage{graphics}

\begin{document}

\begin{center}
{\Large\bf Strong Coupling Expansion for the Pairing Hamiltonian}

\vspace{1cm}

Emil~A.~Yuzbashyan$^{1,2}$, Alexander~A.~Baytin$^{1, 2}$,
and Boris~L.~Altshuler$^{1,2}$

\vspace{0.5cm}

{\small\it

\noindent $^1$ Physics Department, Princeton University, Princeton, NJ 08544\\
 $^2$ NEC Research Institute, 4 Independence Way,
Princeton, NJ 08540\\

}

\end{center}

\begin{abstract}

The paper is devoted to the effects of superconducting pairing in small metallic grains.
It turns out that at strong superconducting coupling and in the limit of large Thouless
conductance one can explicitly determine the low energy spectrum of the problem. We start
with the strong  coupling limit and develop a systematic expansion in powers of the inverse 
coupling constant for the many-particle spectrum of the system. The strong coupling expansion
is based on  the formal exact solution of the Richardson model and converges for realistic values
of the coupling constant. We use this expansion to study the low energy excitations of the system,
in particular energy and spin gaps in the many-body spectrum.

\end{abstract}

\date{}

\section{Introduction}

Since mid 1990's, when Ralph, Black, and Tinkham succeeded in resolving
the discrete
excitation spectrum of nanoscale superconducting metallic grains \cite{RBT}, there has been considerable
 effort  to describe theoretically superconducting correlations in such grains (see e.g. \cite{delft} for
a review). However, very few explicit analytical results relevant
for the low energy physics of superconducting grains have been
obtained, since, in contrast to bulk materials, the discreetness of single electron levels plays an important role.
In this paper we address this problem in the regime of well
developed superconducting correlations.

The electron--electron interactions in weakly disordered grains
with negligible spin--orbit interaction are described by a simple Hamiltonian
\cite{kaa}
\beg
H_{\mbox{univ.}}=H_{BCS}-JS(S+1)
\label{univ}
\en
\beg
H_{BCS}={\sum_{i,\sigma}\eps_i c_{i \sigma }^\dagger c_{i \sigma}-\lam d\sum_{i,
j=1}^{N} c_{i\dn}^\dagger c_{i\up}^\dagger c_{j\up} c_{j\dn}}
\label{BCS}
\en
where $\eps_i$ are single electron energy levels, $d$ is the mean level spacing,
$c_{i \sigma }^\dagger$ and $c_{i \sigma }$ are creation and annihilation
operators
for an electron on level $i$, $S$ and $N$  are
 the total spin and number of
levels respectively.
There are only two sample--dependent coupling constants: $\lam$ and $J$
 that correspond to superconducting correlations and spin--exchange interactions respectively.
 Throughout the present paper, for the sake of brevity, we
 consider only the less trivial case of ferromagnetic exchange,
 $J>0$.

Although Hamiltonian \re{univ} is integrable \cite{integrability1,
integrability2} and
 solvable by Bethe's {\it Ansatz},
the exact solution \cite{theSolution} yields a complicated set of
coupled polynomial equations (see Eq.~(\ref{rich}) bellow). As a
consequence, very few
 explicit results have been derived and
most studies resorted to numerics based on the exact solution. The
purpose of the
present paper is to remedy this situation and to build a simple and intuitive picture of
the low energy
physics of isolated grains in the superconducting phase.

It is well known that  physical observables of a superconductor
are nonanalytic in the coupling constant $\lam$ at $\lam=0$. On
the other hand,  the opposite limit of large $\lam$  turns out to
be regular and relatively simple. Here we use the exact solution to
obtain an explicit expansion in powers of $1/\lam$ for the ground
state and low lying excitation energies.

We will distinguish between two types of excitations:
ones that preserve the number of Cooper pairs (the number of doubly occupied orbitals) and ones
that do not. Only the latter excitations are capable of carrying
nonzero spin. It turns out that for $J=0$ to the lowest order in
$1/\lam$ both types of excitations are gaped with the same gap
$\lam Nd$. We compute explicitly the two gaps to the next nonzero
order in $1/\lam$ and find  the gap for pair--breaking
excitations to be larger. The difference between the two gaps turns out to be of the order of
$d^2/\Delta$, where $d$ is the mean single particle level spacing and $\Delta$ is the BCS energy gap, i.e. the difference vanishes in the thermodynamical limit. We were not able to determine the
convergence criteria for the strong coupling expansion exactly, however we present evidence
that the expansion converges up to realistic values of $\lambda$ between $\lam_{c1}\approx1$
and $\lam_{c2}\approx 1/\pi$.

Hamiltonian (\ref{BCS}) was  studied extensively in 1960's
in the context of pair correlations in nuclear matter (see e.g. \cite{mott}). A straightforward but important observation was
that singly occupied levels  do not
participate in  pair scattering \cite{soloviev}.  Hence, the labels of these levels are good
quantum numbers and their contribution to the total energy is only through the kinetic and
the spin--exchange terms in (\ref{univ}). Due to this  ``blocking effect''
 the problem of diagonalizing the full Hamiltonian (\ref{univ}) reduces to finding
the spectrum of the BCS Hamiltonian (\ref{BCS}) on
the subspace of either empty or doubly  occupied -- ``unblocked''  orbitals.
The latter problem turns out to be solvable \cite{theSolution} by Bethe's {\it Ansatz}.
The spectrum is obtained from the following set of algebraic equations for unknown parameters
$E_i$:
\begin{equation}
-\frac{1}{\lam d}+{\sum_{j=1}^m}\lefteqn{\phantom{\sum}}'\frac{2}{E_i-E_j}=\sum_{k=1}^n\frac{1}{E_i-
2\epsilon_k}\quad i=1,\dots,m
\label{rich}
\end{equation}
where $m$ is the number of pairs and $n$ is the number of unblocked  orbitals
$\epsilon _k$. Bethe's {\it Ansatz} equations  (\ref{rich}) for the BCS Hamiltonian (\ref{BCS})
are commonly referred to as Richardson's equations.
The eigenvalues of the full Hamiltonian (\ref{univ}) are known to be related to Richardson parameters, $E_i$, via
\begin{equation}
E=\sum_{i=1}^{m}
E_i+\sum_B \epsilon_B-JS(S+1)
\label{E}
\end{equation}
where $\sum_{B} \epsilon_{B}$ is a sum over singly occupied -- ``blocked''  orbitals and $S$ is the total spin of blocked orbitals (i.e. the total spin of the system).

BCS results \cite{BCS} for the energy gap, condensation energy,
excitation spectrum, etc.
 are recovered
from  exact solution (\ref{rich}) in the thermodynamical limit \cite{largeN}. The proper limit is
obtained by taking the number of levels, $N$, to infinity, so that
$Nd\to 2D=\mbox{const}$, $\phantom{.}m=n/2=N/2$,
where $D$ is an ultraviolet cutoff usually identified with Debye energy.
In particular, for  equally spaced levels $\eps_i$, the energy gap
$\Delta$ and the ground state energy in the thermodynamical limit are
\beg
\Delta(\lambda)=\frac{D}{\sinh(1/\lambda)}\qquad
E_{\mbox{gr}}^{BCS}=-Dm\coth{1/\lambda}
\label{gap}
\en

Since the BCS Hamiltonian (\ref{BCS})
contains only three energy scales: $D$, $\Delta$, and $d$, there are only two independent
dimensionless parameters: $N$, and $\lambda$. The perturbation theory in small $\lambda$ breaks down in
the superconducting state as is already suggested by BCS formulas (\ref{gap}).
Thus, it is natural to
consider the opposite limit of large $\lambda$ and treat the kinetic term in
Hamiltonian (\ref{univ}) as
a perturbation.

 The paper is organized as follows.  In Section 2
 we consider the limit $\lambda\to\infty$, which is the zeroth
order of our expansion. In this limit one can determine the spectrum straightforwardly
 by representing  the BCS
Hamiltonian
(\ref{BCS})
in terms of Anderson pseudospin operators \cite{ander1}. In particular, one finds that at $J=0$ excitations with nonzero
spin to the lowest order in $1/\lam$ have the same gap (the spin gap) as spinless excitations.
Next, we rederive the same results from Richardson's equations \re{rich} and also show that
in the limit
$\lam \to \infty$  the roots of Richardson's equations are zeroes of Laguerre polynomials.

In Section 3  Bethe's {\it Ansatz} equations (\ref{rich}) are used
to expand the ground state and low--lying excitation energies in series in  $1/\lambda$. We write down
several lowest orders explicitly and give recurrence relations that relate the $k$th order
term to preceding terms. These relations can be used to readily expand up to any
 reasonably high order in $1/\lambda$. Finally, we compute the spin gap to the next nontrivial order in
$1/\lambda$ and demonstrate that at $J=0$
the first excited state always have  zero spin.

\section{The strong  coupling limit}

In this section we analyze  the lowest order of the strong coupling expansion.

As the strength of the coupling constant $\lam$ increases, the
spectrum of the BCS Hamiltonian \re{BCS} undergoes dramatic
changes as compared to the spectrum of noninteracting Hamiltonian
$H_{BCS}(\lam=0)$. First, there is a region of small $\lam$ where the
superconducting coupling causes only small perturbations in the
electronic system. This region shrinks to zero in the
thermodynamical limit and is roughly determined by the condition
$\Delta(\lam)\leq d$ \cite{ander1}, where $\Delta(\lam)$ is given
by \re{gap}. For larger $\lam$ the perturbation theory in $\lam$
breaks down \cite{imry} and strong superconducting correlations
develop in the system. A representative  energy level diagram is
shown on Fig~1. In the crossover regime the spectrum displays
numerous level crossings which reflect the break down of
perturbation theory in $\lam$. The fact that the crossings occur
for random single electron levels $\eps_i$, i.e. in the absence of
any spatial symmetry, is a characteristic feature of quantum
integrability \cite{hubbard}.

 \begin{figure}
\begin{center}
\setlength{\unitlength}{15cm}
\begin{picture}(1, 0.718)(0,0)
   \put(0,0){\resizebox{1\unitlength}{!}{\includegraphics{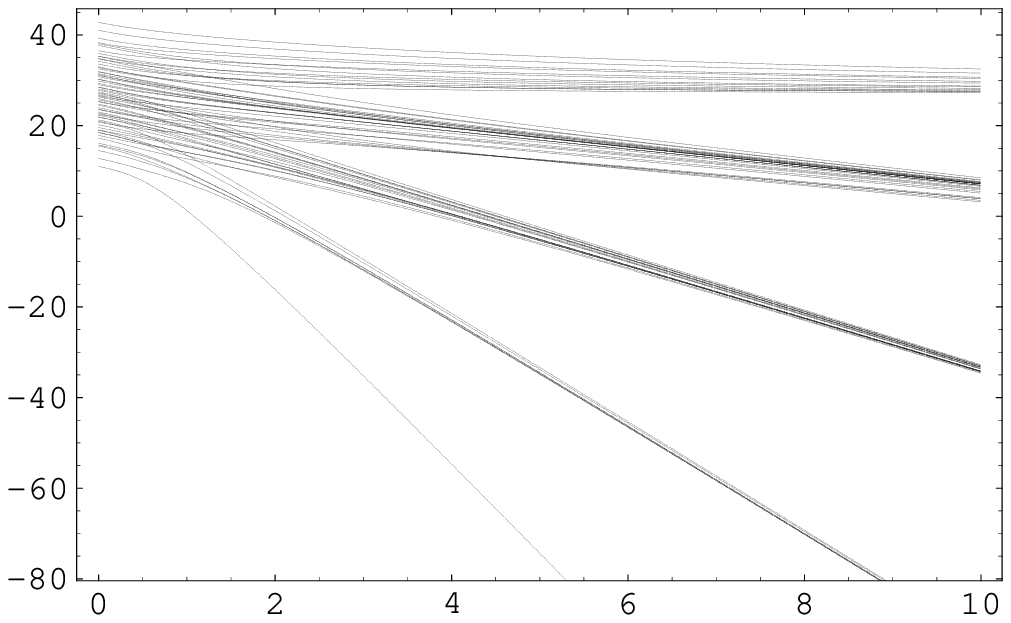}}}
\put(0.085,0.055){\resizebox{0.215\unitlength}{!}{\includegraphics{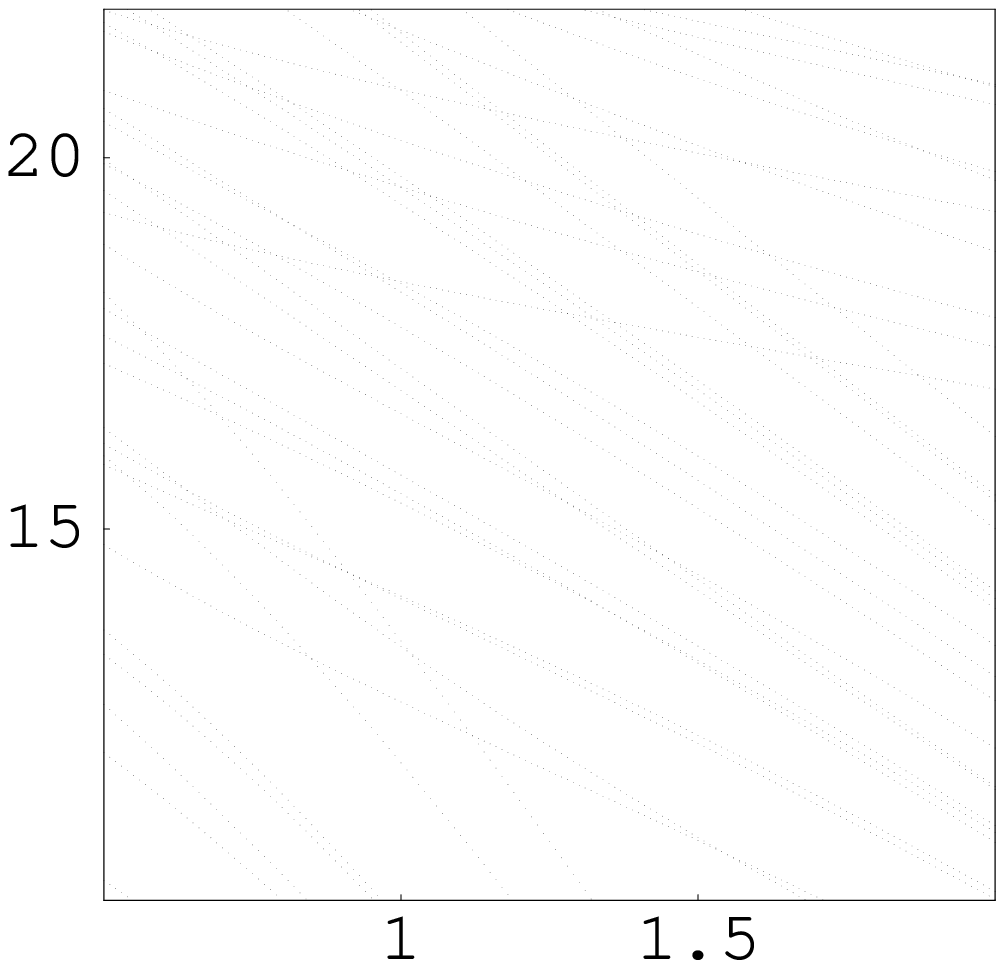}}}
    \put(0.54,-0.03){\makebox(0,0)[b]{\large$\lam$}}
   \put(-0.05,0.35){\rotatebox{90}{\makebox(0,0)[t]{\large Energy}}}
\end{picture}
\end{center}
\caption{Results of exact numerical diagonalization. Energies of  BCS Hamiltonian \re{BCS} for $m=4$ pairs and $n=8$ unblocked single particle levels $\eps_i$ versus coupling constant $\lam$.
All energies are measured in units of the mean level spacing $d$. The single particle levels
$\eps_i$ are computer generated random numbers. As the strength of the coupling $\lam$ increases, the levels coalesce into narrow well separated rays (bands). The width of these bands vanishes in the limit $\lam\to\infty$ (see Eq.~\re{band} and the discussion around it). Slopes of the rays and the number of states in each ray are given by Eq.~(\ref{infiniteh}, \ref{0}, \ref{degeneracy}). The ground state is nondegenerate, while the first group of excited states contains $n-1=7$ states.  Note also the level crossings
for $\lam\sim1$ (see the insert on the above graph). }
\end{figure}

The lowest order of the strong coupling expansion is obtained by neglecting the kinetic energy term in the BCS Hamiltonian \re{BCS}.
This limit can in principle be realized in  a grain
of an ideal regular shape \cite{degenerates}. In this case
 the single electron levels are highly degenerate and if the energy distance between
degenerate many-body levels is much larger than $\g$, only the partially filled Fermi level is
relevant. Then, the
kinetic term in \re{BCS} is simply a constant proportional to the total number of particles
and can be
set to zero.

An efficient way to obtain the spectrum of Hamiltonian \re{univ} in the strong coupling limit
is by representing the interaction term in the
BCS Hamiltonian in terms of Anderson pseudospin-1/2 operators \cite{ander1}.
\begin{equation}
\label{pseudospin}
K^z_i=\frac{c_{i \up }^\dagger c_{i \up}+c_{i \dn }^\dagger c_{i \dn}-1}{2}\qquad
K^+_i=(K^+_i)^{\dagger}=c_{i \uparrow}^{\dagger} c_{i \downarrow}
\end{equation}
The pseudospin is defined only on unblocked levels,
where it has all properties of  spin-1/2, i.e. proper commutation relations and definite value
of $\vec K_i^2=3/4$.

The interaction term in the BCS Hamiltonian \re{BCS} takes a simple form in terms of
$\vec K_i$.

\begin{equation}
\label{infiniteh}
H_{BCS}^{\infty}=-\g \phantom{.}K^+K^-=-\g\left[K(K+1)-(K^z)^2+K^z\right]
\end{equation}
where $\vec K=\sum_i \vec K_i$ is the total pseudospin of the unblocked levels.
The $z$-projection of the total pseudospin  according to
\re{pseudospin} is $K^z=m-n/2$, where $m$ and $n$ are the total
number of pairs and unblocked (either doubly occupied or empty)
levels respectively. It is simple to check that replacing a doubly occupied level
with two singly occupied ones does not affect the difference
$m-n/2$. As a result,
\begin{equation}
\label{kz1}
K^z=m-\frac{n}{2}=M-\frac{N}{2}
\end{equation}
where $M$ is the maximum possible number of pairs and $N$ is the
total number of levels respectively. Hence, the last two terms in
\re{infiniteh} yield a constant independent of the number of
blocked levels. This constant can be set to zero by an overall shift of
all energies. Therefore,  the full Hamiltonian \re{univ}  in the
strong coupling limit is
\begin{equation}
H_{\mbox{univ.}}^\infty=-\g K(K+1)-JS(S+1)
\label{full}
\end{equation}
Since there are $n$ pseudospin-1/2s, the total pseudospin $K$ takes values between $|K^z|$ and
$n/2$,
\beg
\frac{n}{2}\ge K \ge |m-\frac{n}{2}|,
\label{cnstr}
\en
while the total spin $S$ ranges from 0 (1/2) to $M-m$ ($M-m+1/2)$
for even (odd) total number of electrons. For the sake of brevity, let us
from now on consider only the case of even total number of electrons. Then, the sum of the total spin and pseudospin
is constrained by
\beg
K+S\le \frac{N}{2}
\label{constr}
\en
The degree of degeneracy $D(K, S, n)$ of each level is \cite{LL}
\beg
D(K,S)=\frac{n!(2K+1)}{(\frac{n}{2}+K+1)!(\frac{n}{2}-K)!}\phantom{.}
\frac{(N-n)!(2S+1)}{(\frac{N-n}{2}+S+1)!(\frac{N-n}{2}-S)!}
\label{degeneracy}
\en

The ground state of Hamiltonian \re{full} has the maximum possible pseudospin,  $K=N/2$, and
minimal possible spin, $S=0$, provided that $\g>J$  (recall that we consider only positive values
of the exchange coupling $J$).
 
There are two ways to create an elementary excitation. First, one
can decrease the total pseudospin $K$ while keeping the total
number of pairs $M$ unchanged. The second type of excitations
corresponds to breaking pairs and blocking some of the single
electron levels. These excitations can contribute to the total
spin of the grain $S$. They  also affect the pseudospin since its
maximal value $K_{max}=n/2$ is determined by the number of unblocked
levels.  The lowest--lying excitations correspond to $K=N/2-1$,
which can be achieved both with and without breaking a single
Cooper pair. Therefore, we find from \re{full} that the
pair--conserving excitations are separated by a gap
$\Delta_{\mbox{pair}}=N\g$ while pair--breaking excitations can
lower their energy by having nonzero spin $S$. Since the maximum
value of $S$ for two unpaired electrons is $S=1$, we get
$\Delta_{\mbox{spin}}=N\g-2J$.

In the opposite case $J>\g$, $K=0$ and the total spin
has the maximum possible value $S=M$ in the ground state,
i.e. $J=\g$ is the threshold of Stoner instability in the strong coupling limit.

The above results can be obtained directly from  exact solution \re{rich}. Moreover,
individual parameters $E_i$ can also be determined and, since eigenstates
of the BCS Hamiltonian \re{BCS} are given in terms of $E_i$ (see \cite{theSolution}),
this can be used to calculate various correlation functions in the strong coupling limit.

The value of the total pseudospin $K$ turns out to be related to the number, $r$, of those roots
of equations \re{rich} which diverge in the limit $\lam\to\infty$ (see bellow).
To the lowest order in $1/\lam$ we can neglect single electron levels $\eps_i$ in
Eqs.~\re{rich} for these roots
\begin{equation}
-\frac{1}{\g}+{\sum_{j=1}^r}'\frac{2}{E_i-E_j}=
\frac{n'}{E_i}\quad i=1,\dots,r
\label{r}
\end{equation}
where $n'=n+2r-2m$ and summation excludes $j=i$. For the remaining $m-r$ roots we have
\begin{equation}
\sum_{k=1}^n\frac{1}{E_i-2\epsilon_k}=0\quad i=r+1,\dots,m-r
\label{E_m}
\end{equation}
Multiplying each equation in \re{r} by $E_i$ and adding all Eqs.~\re{r},
we obtain the eigenenergies
of the BCS Hamiltonian \re{BCS} for $n$ unblocked levels and $m$ pairs
\begin{equation}
E= -\g\phantom{.} r(n-2m+r+1)
\label{0}
\end{equation}
Comparing this to \re{infiniteh} and \re{kz1}, we find the
relationship between $r$ and $K$
\begin{equation}
r=K+m-n/2 \label{rpseudo}
\end{equation}
Since the total pseudospin, $K$, is constrained by \re{cnstr}, the number, $r$, of diverging Richardson parameters, $E_i$, is also constrained
\beg
\begin{array}{ll}
\dis 2m-n\le r \le m & \mbox{if $n< 2m$}\\
\\
\dis 0\le r \le m & \mbox{if $n\ge2m$}\\
\end{array}
\label{cnstr2}
\en

Bellow in this Section we show that Eqs.~\re{r} have a unique solution. As a result,  the degeneracy
of energy levels \re{degeneracy} is equal to
 the number of solutions
of Eqs.~\re{E_m} for the remaining
 $E_i$. This number can be computed \cite{Gaudin, dukelsky} directly from \re{E_m} and indeed coincides
with \re{degeneracy}.

Finally, Eqs.~\re{r} can be solved to determine parameters $E_i$ to the lowest order in
$1/\lam$ (see  also \cite{s, Sriram}).
To this end it is convenient to introduce a polynomial $f(x)$ of order $r$
with zeroes at $x=x_i=E_i/(\g)$
\begin{equation}
\label{poly}
f(x)=\prod_{i=1}^{r}(x-x_i),
\end{equation}
Using
$$
\lim_{x\to x_i}\frac{f''(x)}{f'(x)}=\sum\limits_{j \ne i} \frac{2}{x_i-x_j}
$$
one can rewrite Eqs.~\re{r} as
\beg
F(x_i)=0\quad\mbox{where}\quad F(x)=x f'(x)-x f''(x)+n'f'(x)
\label{F}
\en
Since $F(x)$ and $f(x)$ are two polynomials of the same degree $r$ with the same roots $x_i$, they
are proportional to each other. The coefficient of proportionality is the ratio of coefficients at
$x^r$ and, according to \re{F}, is equal to $r$. Therefore, $F(x)=rf(x)$, or equivalently
\begin{equation}
\label{lag2}
x f''-(x+n')f'+rf=0
\end{equation}
The only polynomial solution to this equation is  the Laguerre polynomial $L^{-1-n'}_{r}$.
Thus, to the order $\lam$ the nonvanishing roots of Richardson's equations \re{rich} in the
strong coupling limit are determined
by
\begin{equation}
\label{lag3}
L^{-1-n'}_{r}\left(\frac{E_i}{\g}\right)=0\quad n'=n+2r-2m
\end{equation}
where $r$ is the number of nonvanishing roots to the order
$\lam$. This number and the total pseudospin are related by
(\ref{rpseudo}). The ground state has $r=m$, the first degenerate
group of excited states corresponds to $r=m-1$, etc. The constraint $r\ge2m-n$ in \re{cnstr2} follows
from the requirement that the roots of \re{lag3} be nonvanishing \cite{s}. Moreover, it
can be shown \cite{s} using conditions \re{cnstr2} that all Richardson parameters $E_i$ are complex
for even values of $r$, while for odd $r$ there is a single real (negative) root. The fact that the roots of \re{r} are generally complex was also noted  in \cite{dukelsky}  on the basis of numerical solution of Richardson's equations.

\section{The Strong Coupling Expansion}

Now we turn to the expansion in powers of $1/\lam$ around the
strong coupling limit. The evolution of  energy levels with $\lam$
can be viewed as a motion of one--dimensional particles whose
positions are the energies of the BCS Hamiltonian \re{BCS} (see
e.g. \cite{pechukas, hubbard}). Then,  single electron levels
$\eps_i$  determine the initial conditions at $\lam=0$. As the
coupling $\lam$ increases beyond the crossover between the
weakly perturbed Fermi gas and the regime of strong superconducting
correlations, the particles gradually loose the memory of their
initial positions and eventually the spectrum becomes independent
of $\eps_i$. In this limit, the excited levels coalesce into
highly degenerate rays with a universal slope (see Fig.~1 and
Eq.~\re{0}). In the strong coupling expansion the system of
one--dimensional particles evolves from larger to smaller $\lam$.
One expects this evolution to be nonsingular until we come close
to
 the level crossings (see the beginning of the previous section),
i.e. the crossover region, where both expansions
in $\lam$ and in $1/\lam$ break down.

A  quantitative estimate of the convergence of $1/\lam$ expansion can be obtained by considering
various limiting cases. In the thermodynamical limit  the ground state energy is given by BCS expression \re{gap}.  This limit
 is equivalent to keeping only the terms of order $N$
in the $1/\lam$ expansion. We observe from BCS expressions \re{gap} that the expansion in 
$1/\lam$
converges for $\lam>1/\pi$.  In the opposite case of one pair and two levels, $2M=N=2$, the ground state energy can be computed exactly by e.g. solving Eqs.~\re{rich} with the result
\beg
E^2_{\mbox{gr} }=-d(\lam+\sqrt{ 1+\lam^2 } )
\label{2}
\en
In this case the expansion of the ground state energy \re{2} in $1/\lam$ converges for $\lam >1$.
In general, we believe that strong coupling expansion yields convergent rather than asymptotic series with the radius of convergence between $\lam_{c1}\approx 1$ and  
$\lam_{c2}\approx 1/\pi$.

Bellow in this Section we develop an efficient
algorithm for calculating the low energy spectrum to any order in $1/\lam$. While the pseudospin
representation detailed in the previous Section provides a simple and intuitive description of
the strong coupling limit,  the usual perturbation theory becomes unmanageable
beyond the first two orders in $1/\lam$. An approach based on
Bethe's {\it Ansatz} equations, on the other hand, turns out to
be well suited for the purposes of systematic expansion.

\subsection{The ground state}

Here we expand the ground state energy in $1/\lam$. Richardson's equations \re{rich} lead to recurrence relations for the coefficients of the expansion. From these relations the
ground state energy can be computed to any reasonably high order in $1/\lam$, e.g.
 we write down the energy up to $1/\lam^7$.
As it was mentioned above we take the number of electrons to be even and
consider only the  case when $\g>J$. As we have seen in the previous Section,
 this inequality ensures that in the ground state  all levels are unblocked and all electrons are paired, i.e   Richardson's equations \re{rich} should be solved at
$$
m=M\qquad n=N
$$

We begin by introducing a convenient set of variables
\begin{equation}
s_p\equiv\sum_{k=1}^N (2\epsilon_k)^p \qquad
\sigma_p\equiv\sum_{i=1}^M \frac{1}{E_i^p}\label{s1}
\end{equation}
Variables $\sigma_p$ can be expanded into series in the inverse
coupling constant $\lam$.
\begin{equation}
\sigma_p= \sum \limits_{k=0}^\infty a_p^k\lam ^{-k-p} \label{s}
\end{equation}
Next,  we rewrite Richardson's Eqs.~(\ref{rich}) in a form more suitable for our purpose.
We divide the equation for $E_i$ by $E_i^p$ with
$p\ge -1$ and add all $M$ equations for each $p$.
Expanding $1/(1-2\epsilon_k/E_i)$ in $2\eps_k/E_i$ and using an identity
$$
\sum\limits_{i>j}\frac{2}{E_i-E_j}\left(\frac{1}{E_i^p}-\frac{1}{E_j^p}\right)=
p \sigma_{p+1}-\sum\limits_{k=1}^p\sigma_{p-k+1} \sigma_k
$$
we obtain
\beg
\label{eneq1}
\begin{array}{l}
\dis E_{\mbox{gr}}(M, N, s_p)=\sum_{i=1}^M E_i=\\
\dis \qquad\qquad -M(N-M+1)\lam d- \sum_{k=1}^\infty s_k\sigma_k=-M(N-M+1)\lam d-d\sum_{j=0}^\infty
\left(\sum_{k=1}^{j+1}s_ka_k^{j-k+1}\right)\lam^{-j}\\
\end{array}
\en
\beg
\label{peq1}
- \frac{\sigma_p}{\lam d}-\sum_{k=1}^p  \sigma_{p-k+1}  \sigma_k=(N-p)  \sigma_{p+1}
+ \sum_{j=1}^\infty s_j\sigma_{j+p+1}\quad p\ge 0
\en

Now plugging $\sigma_p=\sum_{k=0}^\infty a_p^k\lam ^{-k-p}$ into
the last equation and setting the coefficient at $\lam^{-h-p-1}$
to zero, we obtain \beg \label{maineq}
\frac{a_p^h}{d}+\sum\limits_{k=1}^{p}\sum\limits_{s=0}^h
a_{p-k+1}^{h-s}a_k^s+ \sum\limits_{k=1}^{h}s_k
a_{p+k+1}^{h-k}=-(N-p)a_{p+1}^h \en
Note that from $\sigma_0=M$ it follows
$a^0_0=M$ and $a_0^k=0$ for $k\ge1$. The values of $a^k_0$ serve
as boundary conditions for recurrence relations \re{maineq}. Note
also that according to \re{maineq} the coefficients $a^p_h$ do not
depend on $\lam$ as expected from their definition (\ref{s}). Coefficients $a_p^0$ determine
$\sigma_p$ for the ground state to the lowest nonvanishing order in $1/\lam$ and therefore can be expressed in terms
of zeroes of Laguerre polynomial \re{lag3} with $r=M$. Using \re{lag3}, we obtain
$$
 a_p^0 d^p=(-1)^p\frac{d^p\phantom{x}}{dx^p}\left.\log L^{-1-N}_M(x)\phantom{.}\right|_{x=0}
$$
According to Eq.~(\ref{eneq1}) in order to determine the ground
state energy to order $1/\lam^j$ one has  to calculate the first
$j-p+2$ coefficients $a_p^k$ in the expansion of $\sigma_p$. To do
this, we first
 compute $a^0_p$ for $p\le j+1$, then $a^1_p$ for $p\le j$, then $a^2_p$ for $p\le j-1$, etc.
In other words, we start from $a_1^0$ element of matrix $a_p^h$
and use recurrence relations \re{maineq} to move down the first
column of this matrix until $a^0_{j+1}$, then to move down the
second column from $a^1_1$ to $a^1_j$ etc.

While we were not able to  express  $a_p^k$ in terms of $p$ and
$k$ explicitly, the above procedure allows for  an efficient
calculation, e.g. using {\it Mathematica}, of the ground state
energy to any given order. For example, the ground state energy to
order $1/\lam^2$ is
\beg
\begin{array}{lcl}
\dis E_{\mbox{gr}}(M, N, s_p)
 & = & \dis -M(N-M+1)\lam d+\frac{  s_1M}{N}-\biggl(s_2-\frac{  s_1^2}{N}\biggr)
\frac{M(N-M)}{N^2(N-1)\lam d}\\
\\
&&\dis -\biggl(  s_2-\frac{s_1^2}{N}\biggr)  s_1\frac{M(N-M)(N-2M)}{N^4(N-1)(\lam d)^2}+
\biggl(  s_3-\frac{  s_1  s_2}{N}\biggr)\frac{M(N-M)(N-2M)}{N^3(N-1)(N-2)(\lam d)^2}\\
\end{array}
\label{gr} \en From Eq.~(\ref{gr}) one can make several
observations.

1. For $N=M$ the first two terms give the exact energy. This is seen by noting that $N=M$
means that all levels are doubly occupied, i.e. there is only one state.
Averaging  Hamiltonian \re{univ} over this state gives the exact energy of the system which turns out
to be equal
to the first two terms in \re{gr}.
Therefore, the remaining terms in the $1/\lam$ series for the ground state energy are proportional
to $N-M$.

2. When $N=2M$, all terms with even nonzero powers of $1/\lam$
vanish. This can be demonstrated, e.g., by writing the kinetic
term in the BCS Hamiltonian \re{BCS} in terms of pseudospin
operators \re{pseudospin} \beg H(\lam=0)= \sum_{i=1}^N 2\eps_i
K^z_i+\frac{s_1}{2}\equiv H_0+\frac{s_1}{2} \label{kin} \en
 and noting that $N=2M$ correspond to zero $z$-projection of the total
pseudospin. In this case, by Wigner-Eckart's theorem \cite{angmomentum}, $K_i^z$ has nonzero matrix
elements only for transitions $K\to K\pm1$, while matrix elements for transitions
$K\to K$  are equal to zero.  The terms with even nonzero powers of $1/\lam$ vanish
because they contain at least one matrix element of $H_0$ from \re{kin} between states with the same
$K$.
These terms are therefore proportional
to $N-2M$.
Even terms also vanish when $\eps_i$ are distributed
symmetrically with respect to zero. Hence, they reflect an asymmetry in
the distribution of $\eps_i$. For example, the ground state energy for $N=2M$ and equidistant
single electron levels distributed symmetrically between $\pm D=\pm(m-1/2)d$ is
\begin{equation}
\label{monster}
\begin{array}{c}
E^{2m}_0=-D m
\left [
\dis \lam \frac{2m+2}{2m-1}+\frac{2m+1}{3(2m-1)\lam}-\frac{16 m^2+22 m+7}{180
(2m-1)^2\lam^3}+ \right .
\\
\\
\left .
\dis \frac{128m^3+380m^2+344m+93}{7560(2m-1)^3 \lam^5}+ O(1/\lam^7)
\right ]
\end{array}
\end{equation}
One can check that in the limit  $m\to\infty$ this expression reproduces
 the  BCS result \re{gap} for the ground state energy up to terms of order $1/\lam^7$, while for $m=1$ we recover \re{2}. Note also that the case $p=N$ in Eq.~\re{maineq} does not seem to be problematic as at $p=N$ the factors of $1/(N-p)$ in Eqs.~(\ref{gr}, \ref{monster}) are always compensated by a factor of $(N-p)$ in the numerator
of the corresponding term. 

3. Richardson's equations (\ref{rich}) remain invariant if
 single electron levels $\eps_k$ are shifted by $\delta$
and parameters $E_i$  are shifted by $2 \delta$. The total energy
$E=\sum_{i=1}^M E_i$ then shifts by $2M\delta$. Note that this
shift  is entirely contained in the second term of expansion
(\ref{gr}). Thus, the remaining combinations of $s_k$ at each
power of $1/\lam$ are ``shiftless''. For example,
$$
s_2-\frac{s_1^2}{N}\to s_2+2\delta s_1+N\delta^2- (s_1^2+2N\delta s_1+N^2\delta^2)/N=
s_2-\frac{s_1^2}{N}
$$

\subsection{Excited states}

Let us now expand energies of low--lying excitations  in $1/\lam$.
These expansions turn out to be analogous to that for the ground
state energy.  We begin with  the excitations that conserve the
number of pairs and then turn to the simpler case of
pair--breaking excitations.

It was demonstrated in Section 2 that for $\g>J$ lowest pair--conserving
excitations correspond to total pseudospin $K=N/2-1$ and total spin $S=0$, where $N$ is the total number of single particle levels.  The number of such
states according to degeneracy formula \re{degeneracy} is $N-1$
and their energy is $-\lam dK(K+1)$ according to \re{full}. We
also know from Section 2 that for these states one of parameters
$E_i$ (say $E_M$) remains finite as $\lam\to\infty$, while all
others diverge in this limit.

To distinguish $E_M$ from the rest of parameters $E_i$, we denote it by $\eta$.
 Richardson's equations (\ref{rich}) read
 \beg
-\frac{1}{\g}+\sum_{j=1}^{M-1}\lefteqn{\phantom{\sum}}'\frac{2}{E_i-E_j}=\sum_{k=1}^N
\frac{1}{E_i-2\eps_k}-\frac{2}{E_i-\eta} \quad i<M\label{rest}
\en
\beg
-\frac{1}{\g}-\sum_{j=1}^{M-1}\frac{2}{E_j-\eta}=\sum_{k=1}^N
\frac{1}{\eta-2\eps_k}\quad i=M \label{eta}
\en

Expanding the LHS of Eqs.~(\ref{rest}) in $2\eps_k/E_i$ and
$\eta/E_i$ and performing the same manipulations that lead to
Eqs.~(\ref{eneq1}, \ref{peq1}) for the ground state, we
obtain \beg \sum_{i=1}^{M-1} E_i=-(M-1)(N-M)\lam
d-\sum_{k=1}^\infty (s_k-2\eta^k)\sigma_k \label{exited} \en \beg
- \frac{\sigma_p}{\lam d}-\sum_{k=1}^p  \sigma_{p-k+1}
\sigma_k=(N-p-2)  \sigma_{p+1} + \sum_{j=1}^\infty
(s_j-2\eta^j)\sigma_{j+p+1}\quad p\ge 0 \label{sigmaex} \en where
now $\sigma_p=\sum_{i=1}^{M-1} 1/E_i^p$. We see that replacements
\beg M\to M-1\qquad N\to N-2\qquad s_p\to s_p-2\eta^p \label{repl}
\en
 transform Eqs.~(\ref{exited}, \ref{sigmaex})
into Eqs.~(\ref{eneq1}, \ref{peq1}) for the ground state.
Thus, energies of first $N-1$ excited states are \beg
E_{\mbox{pair}}=\sum_{i=1}^{M-1}E_i+E_M=E_{\mbox{gr}}(M-1, N-2,
s_p-2\eta^p)+\eta \label{ex} \en Let us also rewrite Eq.~\re{eta} for $\eta$ as \beg
\sum_{k=1}^N\frac{1}{\eta-2\eps_k}=-\frac{1}{\g}-2\sigma_1
-2\eta\sigma_2-2\eta^2\sigma_3-\dots \label{etaexp} \en
One can see (by e.g. sketching  the LHS of  Eq.~\re{etaexp} ) that this equation has
$N-1$ roots with the $k$th root lying between $2\eps_k$ and $2\eps_{k+1}$.
To the
lowest order in $1/\lam$ this equation reads \beg
\sum_{k=1}^N\frac{1}{\eta_0-2\eps_k}=0 \label{etamain} \en

Eqs.~\re{sigmaex} and \re{etaexp} are to be solved
iteratively order by order in $1/\lam$. The procedure is similar
to that for the ground state, e.g., recurrent relations analogous
to \re{maineq} can also be derived. The only difference is that
the coefficients at powers of $1/\lam$ now depend also on
$\eta_0$, which has to be obtained from \re{etamain}. For example,
the excitation energies \re{ex} to the first two orders in
$1/\lam$ are \beg E_{\mbox{pair}}=-(M-1)(N-M)\lam
d+\frac{(s_1-2\eta_0)(M-1)}{N-2}+\eta_0 \label{exc} \en \beg
E_{\mbox{pair}}-E_{\mbox{gr}}=N\lam d+\eta_0(1-2f) \label{p} \en
where \beg f=(M-1)/(N-1)\approx M/N\label{fill} \en
 is the filling ratio.

Energies of higher excitations can be computed in the same way by solving
2, 3, 4, \dots coupled equations of the type of (\ref{etamain}). For instance,
energies of the next group of excited levels to the first two orders in $1/\lam$
are determined by solutions of the system
$$
\begin{array}{l}
\dis \sum_{k=1}^n\frac{1}{\eta_1-2\eps_k}=\frac{2}{\eta_1-\eta_2}\\
\\
\dis \sum_{k=1}^n\frac{1}{\eta_2-2\eps_k}=-\frac{2}{\eta_1-\eta_2}\\
\end{array}
$$

Now let us consider  pair--breaking excitations. For $\g>J$ low
energy excitations of this sort correspond to breaking a single
pair of electrons thereby decreasing the number of pairs by 1 and
the number of unblocked levels by 2. Let  the single electron
levels occupied by two unpaired electrons have energies $\eps_a$
and $\eps_b$. Since the lowest energy is achieved by having the
unpaired electrons in a triplet state (recall that $J>0$ corresponds to the ferromagnetic exchange), the energy of lowest
pair--breaking excitations  according to \re{E} is
\begin{equation}
E_{\mbox{spin}}=\eps_a+\eps_b-2J+E_{\mbox{gr}}(N-2, M-1, s_p-(2\eps_a)^p-(2\eps_b)^p)
\label{enspin}
\end{equation}
Note that, unlike $\eta$ in \re{ex}, single electron energies $\eps_a$ and $\eps_b$ do not depend on
$\lam$. Therefore, to compute the energy of pair--breaking excitations we  need only
recursion relations \re{maineq} for the ground state with $N'=N-2$, $M'=M-1$, and
$s'_p=s_p-\eps_a^p-\eps_b^p$.
In particular, to the first two orders in $1/\lam$ we get from \re{gr}
$$
E_{\mbox{spin}}=-(M-1)(N-M)g+\frac{(s_1-2\eps_a-2\eps_b)(M-1)}{N-2}+\eps_a+\eps_b-2J
$$
\beg
E_{\mbox{spin}}-E_{\mbox{gr}}\approx N\lam d+(\eps_a+\eps_b) (1-2f)-2J
\label{un}
\en

It is instructive to compare the above results with  the BCS theory \cite{BCS}.  For this purpose
let us write down the energies of the pair--conserving excitations for large $M$ and $N$
up to the order $1/\lam$.
\beg
 E_{\mbox{pair}}-E_{\mbox{gr}}\approx 2D\lam+\eta_k(1-2f)+\eta_k^2 \frac{f(1-f)}{D\lam}
\label{ed} \en where $D=Nd$, $f$ is the filling ratio \re{fill},
and $\eta_k$ is the $k$th root of Eq.~\re{etaexp}. In
deriving the above equation from \re{p} and \re{gr} we shifted the
single electron levels so that $\bar
\eps_i=\left(\sum_{i=1}^N\eps_i\right)/N=0$. In BCS theory (i.e.
in the  limit $N, M\to\infty$) pair--conserving excitation energies
are \cite{largeN} \beg 2\sqrt{ (\eps_k-\mu)^2 +\Delta^2}
\label{exbcs} \en where $\mu$ is the chemical potential and
$\Delta$ is the gap. In the strong coupling regime both $\mu$ and
$\Delta$ are of order $\lam$. Expanding the square root in
expression \re{exbcs} in small $\eps_k$ up to $\eps_k^2$, we see
that \re{exbcs} and \re{ed} coincide to this order if we identify
$$
\eta_k=2\eps_k \quad \Delta =D\lam \sqrt{ 4f(1-f)}\quad \mu=(2f-1)D\lam
$$
The first of these equations follows from \re{etaexp} in the limit of large $N$, while the remaining
two  can be derived from the BCS equation for the gap and chemical potential (see e.g.
\cite{largeN}).  Similarly one can check that pair--breaking excitations \re{enspin} correspond to two Bogoliubov  quasi-particles with   total energy
\beg
\sqrt{(\eps_a-\mu)^2+\Delta^2}+\sqrt{(\eps_b-\mu)^2+\Delta^2}
\label{ed1}
\en
 Note that in the BCS limit the difference between pair--breaking and pair--conserving excitations disappears and expression \re{exbcs}  simply corresponds to two quasi-particles
in a singlet state each having the energy $\sqrt{ (\eps_k-\mu)^2 +\Delta^2}$.

We have seen in Section 2 (see also Fig.~1) that in the strong coupling limit many-particle energy levels of the BCS Hamiltonian \re{BCS} coalesce into narrow well separated bands. Expression \re{ed} can be used to estimate the ratio of  the width of the first band, $W_1$, to the single particle bandwidth $D=Nd$.
\beg
\frac{W_1}{D}\approx 2(1-2f)+\frac{f(1-f)}{\lam}
\label{band}
\en
where $W_1$ is the width. Note that at half filling, $f=1/2$, the width of the first band goes to zero
as $\lam\to\infty$.  In general, it follows from  Wigner-Eckart's theorem \cite{angmomentum} (see the discussion in item 2 under the ground state formula \re{gr}) that at half filling widths of higher bands
also vanish as $\lam\to \infty$.

According to the  BCS equations for the excitation energies \re{ed} and \re{ed1}  the gaps
$\Delta_{\mbox{spin}}=\left[E_{\mbox{spin}}-E_{\mbox{gr}}\right]_{min}$
and
$\Delta_{\mbox{pair}}=\left[E_{\mbox{pair}}-E_{\mbox{gr}}\right]_{min}$
 for the two types of excitations coincide in the thermodynamical limit.
We have also seen in Section 2 (see the discussion bellow degeneracy
formula \re{degeneracy}) that when $0<J<\lam d$ and $J/(\lam d)$
remains finite as $\lam\to\infty$,
 spin-1 excitations have lower energy as compared to pair--conserving excitations
. If, however, $J\sim d$ or smaller, keeping $J$ to the lowest
order in $1/\lam$ in excitation energy \re{enspin} is not
justified. In this case the two gaps  are the same to this order.
Therefore, it is interesting to set $J=0$ and evaluate the gaps to
the next nonzero order.

Depending on the filling ratio $f$ (see Eq.~(\ref{fill})) we can
distinguish two different cases.

1. $f\ne1/2$. Lowest lying excitations correspond to smallest or largest possible values of $\eta_0$ and
$\eps_a+\eps_b$ depending on the sign of $(1-2f)$. To determine the maximal and minimal $\eta_0$,
note that the $k$th root of
Eq.~\re{etamain} lies between $2\eps_k$ and
$2\eps_{k+1}$. If
$N$ is large and $\eps_k-\eps_{k+1}\to 0$ as $N\to\infty$, the smallest and
largest solutions of (\ref{etamain}) are $\eta_0^{min}\approx 2\eps_1$ and
$\eta_0^{max}\approx 2\eps_n$ respectively.
 We have from (\ref{p}, \ref{un})
\beg
\Delta_{\mbox{spin}}-\Delta_{\mbox{pair}}=d|1-2f|>0
\label{bellow}
\en
where we have used $\eps_n-\eps_{n-1}\approx \eps_2-\eps_1\approx d$ and $d$ is the mean level spacing. 

2. $f=1/2$. To the first two orders in $1/\lam$:
$\Delta_{\mbox{spin}}-\Delta_{\mbox{pair}}=0$. In the next order we obtain from
(\ref{gr}, \ref{ex}, and \ref{enspin})
\beg
\Delta_{\mbox{pair}} - \Delta_{\mbox{spin}}= \frac{\eta_0^2-2\eps_a^2-2\eps_b^2}{2N\lam d}
\label{half}
\en
where we shifted single electron levels so that $\bar \eps_i=\left(\sum_{i=1}^N\eps_i\right)/N=0$.
We show in the Appendix using  Eq.~\re{etamain} for $\eta_0$ that the minimal value
of $\eta_0^2$ is always smaller than that of $2(\eps_a^2+\eps_b^2)$.
Therefore,
$\Delta_{\mbox{spin}}>\Delta_{\mbox{pair}}$.

Thus, at $J=0$ the pair--breaking  excitations always have a larger gap in the strong coupling limit.
Note that for $\lam=0$ the situation is opposite as it always costs less energy to move one of the two
electrons on the highest occupied single electron levels to the next available level. Since according to BCS expression \re{gap} the energy gap in the strong coupling limit is $2\Delta\approx 2D\lam = N\lam d$, we see from \re{half} that at the half--filling 
$\Delta_{\mbox{pair}} - \Delta_{\mbox{spin}}\approx d^2/\Delta$, i.e. the difference between the two gaps vanishes in the thermodynamical limit.  

\section{Conclusion}

We determined the spectrum of the Universal Hamiltonian \re{univ}
in the strong superconducting coupling ($\lam \geq 1$) limit
(\ref{full}, \ref{degeneracy}, \ref{lag3}) and developed a
systematic expansion in $1/\lam$ around this limit (\ref{maineq},
\ref{gr}, \ref{ex}, \ref{enspin}) for the ground state and
low--lying excitation energies. We detailed an algorithm by which
these energies can be explicitly evaluated up to arbitrary high order in $1/\lam$ and estimated that the expansion
converges for $\lam>\lam_c$ where $\lam_c$ lies between $\lam_{c1}\approx1$
and $\lam_{c2}\approx 1/\pi$.
Technically, this expansion is based on the existence of the exact solution
\cite{theSolution} of the BCS Hamiltonian \re{BCS}. We found that in the strong coupling limit
Richardson parameters are zeroes of appropriate Laguerre polynomials \re{lag3} and analyzed
their behavior at large enough but finite $\lam$  .

We found that it is important to distinguish between two types of
excitations in the problem: those that conserve the total number
of paired electrons and those that do not. We determined the
energy gaps for both types and found that at zero spin--exchange
constant, $J=0$, in contrast to the weak superconducting coupling
limit, the gap for pair--breaking excitations is always larger
(\ref{bellow}, \ref{half}).

We believe there are two physically motivated questions within the scope of validity
(see \cite{kaa}) of the Universal
Hamiltonian \re{univ} that still
need further clarification.
The first problem is to develop a quantitative  description
of the crossover between a perturbed Fermi gas and the region of strong superconducting correlations
(see \cite{imry} and the discussion in the beginning of Sections 2 and 3).
The second problem is to study analytically
the interplay between  superconducting correlations and spin--exchange
(see e.g. \cite{fazio}).

\section{Appendix}

We show here using  Eq.~\re{etamain} for $\eta_0$ that the minimal value
of $\eta_0^2$ is always smaller than that of $2(\eps_a^2+\eps_b^2)$, i.e.
\beg
x_0^2 < 2(a^2+b^2)
\label{ineq}
\en
where $x_0$ is the smallest in absolute value solution of \re{etamain}, $a$ and $b$ are the two smallest in absolute
value single electron levels $\eps_i$, and  $|a|\leq  |b|$.
Indeed, consider a function
\beg
g(x)=\sum_{k=1}^N\frac{1}{x-2\eps_k}
\label{gx}
\en
To prove \re{ineq} we need to show that $g(x)$ has a zero on the interval $(-c, c)$, where
$$
c=\sqrt {2( a^2+b^2)}
$$
For $N=2$ there is only one zero, $x_0=\eps_1+\eps_2$, and \re{ineq} clearly holds. Consider $N>2$.
\vspace{0.3cm}

\begin{figure}[h!]
\begin{center}
\setlength{\unitlength}{7cm}
\begin{picture}(1, 0.718)(0,0)
   \put(0,0){\resizebox{1\unitlength}{!}{\includegraphics{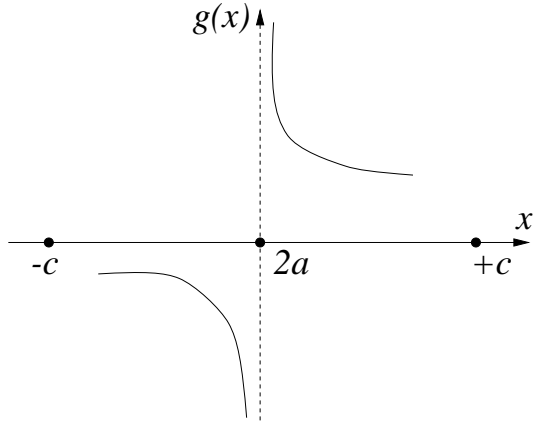}}}
\end{picture}
\end{center}
\caption{ A schematic plot of the function $g(x)=\sum_{k=1}^N \frac{1}{x-2\eps_k}$ on the interval from $-c$ to $c$, where $c=\sqrt {2( a^2+b^2)}$,    $a$ and $b$ are the two smallest in absolute
value single electron levels $\eps_i$, and  $|a|\leq  |b|$. Note that since $2|b|>c$ there is only one
pole on this interval. In the vicinity of $2a$ we have $g(x)\approx 1/(x-2a)$ and therefore $g(x)$ is positive on the immediate right of $x=2a$ and negative on the left.}
\end{figure}

First, note that  $g(x)$ has a single pole at $x=a$ on this interval from $-c$ to $c$, and $g(a+)>0$, while $g(a-)<0$.
Hence, there is a zero between
$c$ and $-c$ iff either $g(c)<0$ or $g(-c)>0$. To show that this is the case it is sufficient to demonstrate that
$g(c)-g(-c)<0$. We have
$$
g(c)-g(-c)=\sum_{i=1}^N \frac{2c}{c^2-4\eps_i^2}=\sum_{\eps_i\ne a, b} \frac{2c}{c^2-4\eps_i^2}
$$
which is indeed negative since $c^2<4\eps_i^2$ for all $\eps_i$ except $\eps_i=a, b$.


\begin{thebibliography}{99}

\bibitem{RBT} D.C. Ralph, C.T. Black and M. Tinkham: {\it Phys. Rev. Lett.}
{\bf 76}, 688 (1996); {\bf 78},
4087 (1997).

\bibitem{delft} J. von Delft: {\it Annalen der Physik}
(Leipzig), 10, {\bf 3}, 219-276 (2001);\\
{\tt cond-mat/0101021}

\bibitem{kaa} I.L. Kurland, I.L. Aleiner, B.L. Altshuler:
{\it Phys. Rev. B} {\bf 62}, 14886 (2000);\\
{\tt cond-mat/0004205}

\bibitem{integrability1} L. Amico, A. Di Lorenzo, A. Osterloh: {\it Phys. Rev. Lett.}
{\bf 86}, 5759 (2001)

\bibitem{integrability2} M.C. Cambiaggio, A.M.F. Rivas, M. Saraceno: {\it Nucl. Phys. A} {\bf 424},
157 (1997)


\bibitem{theSolution} R.W. Richardson and N. Sherman: {\it Nucl. Phys.} {\bf 52},
221 (1964); {\bf 52}, 253 (1964).

\bibitem{mott} Bohr A. and Mottelson B. R.: {\it Nuclear Structure}, W. A. Benjamin, New York,
(1969)

\bibitem{soloviev} V. G. Soloviev: {\it Mat. Fys. Skrif. Kong. Dan. Vid. Selsk.} {\bf 1} (1961) 1

\bibitem{BCS} J. Bardeen, L.N. Cooper, J.R. Schriefer: {\it Phys. Rev.} {\bf 108} 1175 (1957)

\bibitem{largeN} R.W. Richardson: {\it J. Math. Phys.} {\bf 18},1802 (1977).

\bibitem{ander1} P.W. Anderson: {\it Phys. Rev.} {\bf 112}, 1900 (1958).


\bibitem{ander} P. W. Anderson: {\it J. Chem. Solids} {\bf 11}, 26 (1959)

\bibitem{imry}
 M. Schechter, Y. Imry, Y. Levinson, J. von Delft: {\it Phys. Rev. B}
{\bf 63}, 214518 (2001)

\bibitem{hubbard} E. A. Yuzbashyan, B. L. Altshuler, B. S. Shastry:
{\it J. Phys. A} {\bf 35}  34  7525-7547 (2002) {\tt  cond-mat/0201551}

\bibitem{degenerates} O. Bozat , Z. Gedik: {\tt cond-mat/0110318}

\bibitem{LL}
Landau L. D. and Lifshitz E. M. : {\it Quantum Mechanics}, Pergamon Press, Oxford,
\S 63, Problem 1, pp. 239-240, (1991)

\bibitem{Gaudin}
M. Gaudin: {\it Mod\`eles exactament r\'esolus}, Les \'Editions de Physique, France, (1995).

\bibitem{dukelsky} J. M. Roman, G. Sierra, J. Dukelsky: {\it Phys. Rev. B} {\bf 67}, 064510 (2003).

\bibitem{s} G. Szeg\"o: {\it Orthogonal Polynomials},  AMS, New York (1939).  On p. 138 Szeg\"o shows that zeroes of an appropriate Laguerre polynomial obey Eqs.~\re{r}.  Theorem 6.37 on p. 147. gives  the number of real positive and negative roots of  a Laguerre polynomial $L^\alpha_n(z)$ in terms of $n$ and $\alpha$.  

\bibitem{Sriram}  B. S. Shastry and  A. Dhar:
{\it J. Phys. A} {\bf 34}  31 6197-6208 (2001) {\tt cond-mat/0101464}




\bibitem{pechukas}
Pechukas P. 1983 {\it Phys. Rev. Lett.} {\bf 51} 943

\bibitem{futureN} E.A. Yuzbashyan, A.A. Baytin, B.L. Altshuler: to appear.

\bibitem{angmomentum} A.R. Edmonds: {\it Angular momentum in quantum mechanics} ,
 2nd ed.,  Princeton University Press, c1960.

\bibitem{fazio}  G. Falci, Rosario Fazio, A. Mastellone: {\tt cond-mat/0208259}

\end{thebibliography}
\end{document}